\documentclass[aip,jap,reprint,floatfix,superscriptaddress,amsfonts,amsmath,amssymb]{revtex4-2}
\usepackage{graphicx}

\begin{document}

\title{Electrical conductivity of conductive films based on random metallic nanowire networks} 

\author{Yuri Yu. Tarasevich}
\email[Corresponding author: ]{tarasevich@asu-edu.ru}
\affiliation{Laboratory of Mathematical Modeling, Astrakhan Tatishchev State University, Astrakhan, Russia}

\author{Andrei V. Eserkepov}
\email{dantealigjery49@gmail.com}
\affiliation{Laboratory of Mathematical Modeling, Astrakhan Tatishchev State University, Astrakhan, Russia}

\author{Irina V. Vodolazskaya}
\email{irina.vodolazskaya@asu.edu.ru}
\affiliation{Laboratory of Mathematical Modeling, Astrakhan Tatishchev State University, Astrakhan, Russia}

\date{\today}

\begin{abstract}
Using computer simulation, we investigated the dependence of the electrical conductivity of random two-dimensional systems of straight nanowires on the main parameters. Both the resistance of the conductors and the resistance of the contacts between them were taken into account. The dependence of the resistance, $R$, between network nodes on the distance between nodes, $r$, is $R(r) = R_\Box/\pi \ln r + \mathrm{const}$, where $R_\Box$ is the sheet resistance.
\end{abstract}

\maketitle

\section{Introduction\label{sec:intro}}
Interest in studying the conductive properties of transparent conductive films (TCFs) based on random nanowire networks (RNNs) is due to their numerous technological applications \cite{Langley2013,Sannicolo2016,Papanastasiou2020,Ding2024}. Although the sheet resistance of such networks can be calculated directly, the calculations require many characteristics of the system (distributions of lengths, diameters and resistances of nanowires, distribution of junction resistance), the measurement of which is difficult. Furthermore, such calculations can hardly offer an analytical dependence of the sheet resistance on the basic physical parameters of the systems under consideration. Although various theoretical approaches offer such analytical dependencies, they are often based on more or less reasonable assumptions rather than rigorously proven statements.

\citet{OCallaghan2016} studied the RNNs accounting for both wire resistance and the contact resistance. They applied the effective medium theory to find the dependency of the sheet resistance of TCFs on the concentration of nanowires. Several assumptions were stated and used in their consideration, e.g., that dense RNNs can be considered as square lattices with conductive edges. \citet{OCallaghan2016} demonstrated that, for both RNNs and a square lattice, dependencies of the two-point resistances on the distance between nodes are similar. This observation was used to replace an RNN by a square lattice \cite{OCallaghan2016}. In fact, this similarity only confirms the obvious fact that a TCF based on randomly distributed metallic nanowires is a conductive plane. Indeed, for two-dimensional systems, the asymptotic dependence of the resistance between two points on the distance between these points is expected to be logarithmic.

\citet{Venezian1994} derived the resistance between two arbitrary nodes of a square grid of identical resistors. The resistance, $R_{l,m}$ between nodes $(0,0)$ and $(l,m)$ is
\begin{equation}\label{eq:Venezian}
\frac{R_{l,m}}{R_0}= \frac{1}{\pi} \ln\sqrt{ l^2 + m^2 } + 0.51469,
\end{equation}
when $l^2 + m^2$ is large enough. Here and hereinafter, the lattice constant is unit, $R_0$ is the resistance of one resistor in the grid. Equation~\eqref{eq:Venezian}  differs slightly from that presented in the original work, since we have corrected a misprint and performed an equivalent transformation to easier comparison with succeeding formulas. \citet{Venezian1994} claimed that the array of resistors can be treated as a lumped-parameter model for a conductive plane of conductivity $\sigma$ and thickness $d$. Using the square grid, \citet{Venezian1994}  showed that the lumped-parameter model of a plane with a sheet resistance equal to $R$ is a grid of equal resistors of resistance $R$. The two-point resistance of the conductive plane is
\begin{equation}\label{eq:Venezianplane}
R(r) = \frac{R}{\pi} \ln r +\mathrm{const}.
\end{equation}
$R(r)$ is the resistance between two points at distance $r$, $R$ is the sheet resistance of the conductive plane; the dimensional constant has to have an appropriate dimensionality \cite{Mamode2019}.

Similar formula have been derived using Green's function method \cite{Cserti2000,OCallaghan2016}
\begin{equation}\label{eq:R2OCallaghan}
R(r) \approx \frac{R_0}{\pi} \left(\ln r + \gamma + \frac{\ln 8}{2}\right),
\end{equation}
where $\gamma \approx 0.57721$ is the Euler--Mascheroni constant; $\mathbf{r} = l \mathbf{a}_1 + m\mathbf{a}_2$, $\mathbf{a}_i$ are the primitive vectors of the square lattice; since $|\mathbf{a}_i|=1$, hence, $r = \sqrt{l^2 + m^2}$.  When $r=1$, the rest constants have to correspond to the two point resistance between the nearest nodes of the lattice; this resistance is $R_0/2$ for the square network.

\citet{Melnikov2018} have studied a square lattice of resistors using both an analytical consideration and a computer simulation. The conductance of the resistors was chosen according to the truncated Gaussian distribution with the mean value $g_0$ and standard deviation $0.2g_0$. In this case, the effective resistance is $R_\text{eff} = 1.021 g_0^{-1}$. The theoretical prediction of dependency of the two-point resistance on the distance between nodes was in good agreement with the direct computations as well as with the analytical results \cite{Venezian1994}.

Similar asymptotic dependence of the two-point resistance on the distance between the points was obtained for the triangular lattice\cite{Owaidat2018}
\begin{equation}\label{eq:Rtriangular}
 R(r) = \frac{R_0}{\pi \sqrt{3}} \left(\ln r + \gamma + \frac{1}{2} \ln 12\right).
\end{equation}
Note, that $R/\sqrt{3}$ is the electrical resistivity of the infinity triangular lattice.

Using a Y--$\Delta$ transformation, a hexagonal lattice can be represented as a triangular lattice, from which it follows that
\begin{equation}\label{eq:R2hex}
  R(r) = \frac{\sqrt{3}R_0}{\pi}\left( \ln r + \gamma + \ln 2 \right).
\end{equation}
Again,  $R_0\sqrt{3}$ is the electrical resistivity of the infinity honeycomb lattice.

Thus, from general electrostatic reasons, the asymptotic behavior of the average two-point resistance on the distance is expected to be
\begin{equation}\label{eq:R2asymp}
  R(r) \approx \frac{R_\Box}{\pi} \ln r + \mathrm{const}
\end{equation}
for any planar, dense, isotropic, and homogeneous resistor network, regardless of whether it is regular or irregular, uniform or random. Here, $R_\Box$ is the equivalent sheet resistance of such a resistor network. It would be proportional to the branch resistance $R_0$ rather than exactly equal to it, as in the case of the square grid of equivalent resistors studied by \citet{Venezian1994}.

Our goal is a study of the resistance between two nodes of RNNs. The rest of the paper is constructed as follows. Section~\ref{sec:methods} provides necessary information and describes some technical details of simulation. Section~\ref{sec:results} presents  our main findings. Section~\ref{sec:concl} summarizes the main results.

\section{Methods and preliminary results\label{sec:methods}}
\subsection{Some basic information}
The nanowires are often mimicked as zero-width conductive segments (sticks) \cite{Yi2004,Kumar2016,OCallaghan2016,Kim2018,Ainsworth2018}, since the typical ratio of nanowire length to its diameter is 100 of order of magnitude \cite{Tarasevich2023a}. Although model of zero-width sticks is widely used, it has some obvious limitations. Since our focus are nanowires with large but finite aspect ratio, in our 2D model, the distance between the two nearest contacts has to be larger than the nanowire diameter. Effect of finite wire width on the network properties was recently studied \cite{Daniels2021}.

Let there be a domain $L \times L$ with periodic boundary conditions (PBCs). Let there be $N$ zero-width line segments (sticks) of equal length $l$ ($l \ll L$). These sticks are deposited in such a way that coordinates of their centers are independent and identically distributed within the domain, while their orientations are equiprobable.

When the number density of sticks exceeds the percolation threshold, the RNN can carry the electrical current. Far above the percolation threshold, almost all sticks belong to the percolation cluster, while almost all segments in the percolation cluster belong to its backbone, i.e., the current carrying fraction of the RNN \cite{Kumar2017,Tarasevich2021}.

\begin{figure}[!htb]
  \centering
  \includegraphics[width=\columnwidth]{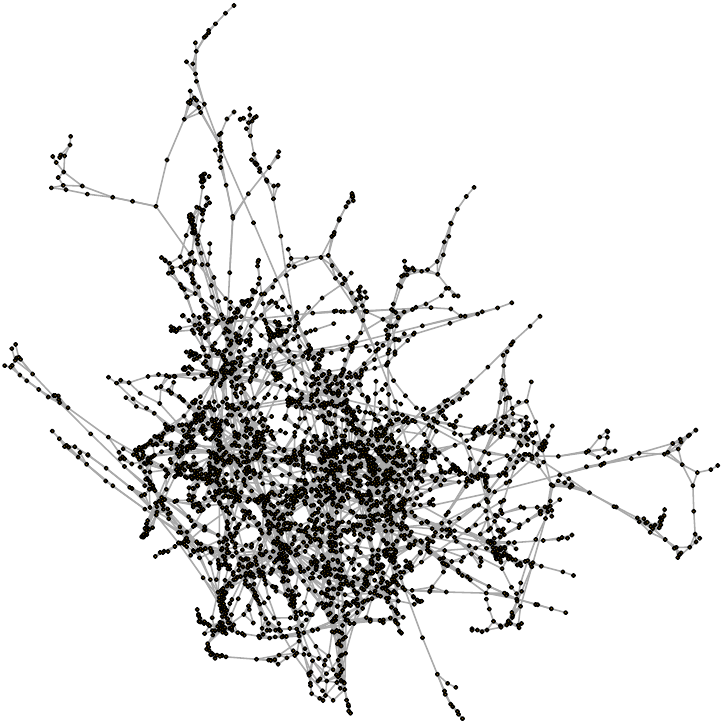}
  \caption{Example of a network derived from RNN when only junction resistances of this RNN are taken into account.\label{fig:networkJDR}}
\end{figure}

There are three distinct possibilities.
\begin{itemize}
  \item When the junction resistance, $R_j$, in a RNN is negligible as compared to the wire resistance, $R_w$ ($R_w \gg R_j$), the RNN is a planar network and tends to a 4-regular network  as the number density of wires increases, while the resistances of edges are random variables which values are proportional to edge lengths.
  \item When the junction resistance and the wire resistance are of the same order, as the number density of sticks increases, the RRN tends to a nonplanar 3-regular network with randomly distributed edge resistances.
  \item When a wire resistance in a RNN is negligible as compared to the junction resistance, $R_w \ll R_j$, the RNN is a nonplanar irregular network with equal edge resistances, $R_j$  (Fig.~\ref{fig:networkJDR}). The degree of nodes corresponds to the Poisson distribution
      \begin{equation}\label{eq:MPoiss}
 f(k; \lambda)  = \frac{\lambda^k e^{-\lambda}}{k!},
\end{equation}
where
\begin{equation}\label{eq:lambda}
\lambda = \frac{2 n l^2}{\pi}
\end{equation}
is the mean (see, e.g., Ref.~\onlinecite{Ainsworth2018}). Here,
\begin{equation}\label{eq:n}
  n = \frac{N}{L^2}
\end{equation}
is the number density of sticks.
\end{itemize}

\subsection{Two-point resistance of some regular networks}
To check our software, we compute the resistance between two nodes of a regular lattice. We used a square lattice of size $L=128$, while edge length, $l$, and edge resistance, $R_0$, were unit. One node was taken near the lattice center; its coordinates were assumed to be $(0,0)$. Applying Ohm's law to each resistor and Kirchhoff's point rule to each junction, a system of linear equations was obtained. This system was solved numerically.

\begin{figure}[!htb]
  \centering
  \includegraphics[width=\columnwidth]{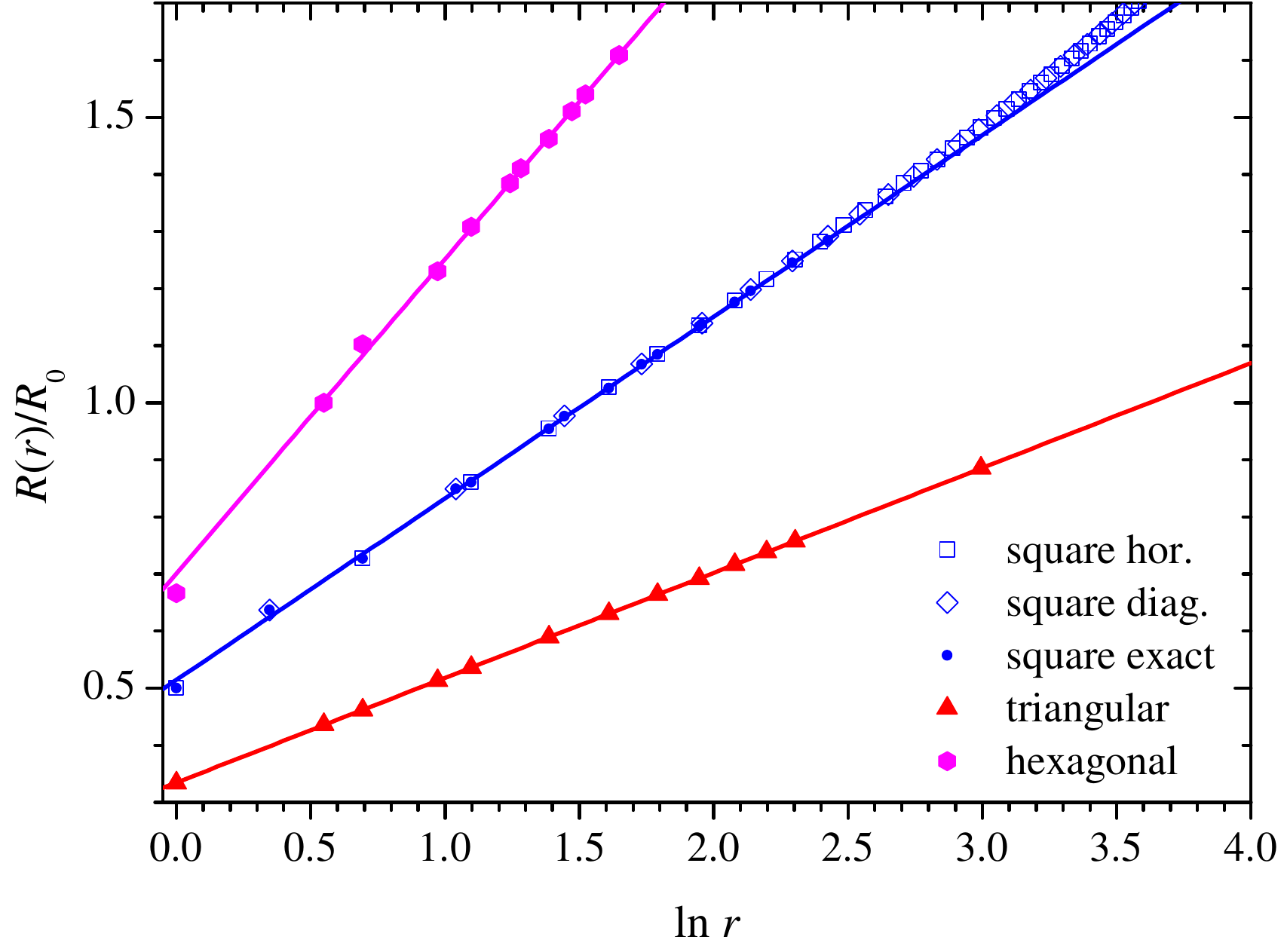}
  \caption{Resistance between two nodes against the distance between these nodes. Comparison of direct computations (squares and diamonds) with a theoretical predictions~\eqref{eq:Venezian} (solid line) and exact values\cite{Flanders1972} (points) for a square lattice; as well as exact values for a triangular lattice  (triangles) and honeycomb lattice (hexagons)\cite{Atkinson1999} along with corresponding asymptotic behavior~\eqref{eq:Rtriangular} and~\eqref{eq:R2hex}.}\label{fig:R2pointsquaretrian}
\end{figure}

Figure~\ref{fig:R2pointsquaretrian} compares the theoretically predicted behaviour  of the resistance between two nodes of the square lattice~\eqref{eq:Venezian} with computed values. Here, empty squares correspond to nodes located to right from the reference node, their coordinates are $(0,i)$, while diamonds correspond to nodes located on the diagonal, their coordinates are $(i,i)$.  Since~\eqref{eq:Venezian} is valid for an infinite grid, while our computations were performed for a lattice of a finite size, a boundary effect is expected. Figure~\ref{fig:R2pointsquaretrian} suggests that boundary effect is negligible up to the distance from the lattice center approximately $L/4$. We consider Fig.~\ref{fig:R2pointsquaretrian} as a validation of our software for the computations of the resistance between two points.

Besides, Fig.~\ref{fig:R2pointsquaretrian} exhibits behavior of the resistance between two nodes for triangular and honeycomb lattices. Theoretically predicted values of the resistance between two nodes on the triangular lattice\cite{Atkinson1999} (shown as triangles) are fairly close to the asymptotic line~\eqref{eq:Rtriangular} even for small values of $r$.
Theoretically predicted values of the resistance between two nodes on the honeycomb lattice\cite{Atkinson1999} (shown as hexagons) are also fairly close to the asymptotic line~\eqref{eq:R2hex}.

The linear dependence of the resistance between nodes on the logarithm of the distance between these nodes is also expected for several other lattices, since these lattices can be transformed into a square or a triangular lattice \cite{Cserti2011,Owaidat2018,Owaidat2021}. Thus, the behavior of the resistance between two points on above mentioned lattices supports our proposition that the resistance between  two nodes for \emph{any homogeneous and isotropic dense 2D system} is expected to be proportional to the logarithm of the distance between these nodes, while the coefficient of the proportionality for each particular system is the sheet resistance divided by $\pi$.

To produce a sample of the RNN, we used Newman--Ziff algorithm \cite{Newman2000,Newman2001}, viz., sticks of unit length ($l=1$) were randomly deposited one by one within the domain until the percolation threshold, $n_c$. In our simulation $n_c = 5.53 \pm 0.11$. Then the number density of sticks was successively increased to $n_c + 5$, $n_c + 10$, $n_c + 15$ and so on.

To compute the resistance between two nodes in the case of RNNs, we used systems of size $L=32$. We computed the resistance between a node located near the system center and all the rest nodes located within a ring of radius $r$ and width $dr=0.1$. The resistances were averaged for each ring. For each value of the number density of wires, 10 samples of RNNs were generated. The results were averaged over these 10 samples. We used the same set of systems for all three cases. For the case, when the wire resistance and the junction resistance are of equal importance, we set $R_w=R_j=1$ arb. units. For the case, when the junction resistance dominates over the wire resistance, we set $R_j = 1$, $R_w = 10^{-6}$ arb. units. For the case, when the wire resistance dominates over the junction resistance, we set $R_w = 1$, $R_j = 10^{-6}$ arb. units.  The error bars in the figures correspond to the standard deviation of the mean. When not shown explicitly, they are of the order of the marker size.

To calculate the sheet resistance, we attached a pair of superconducting buses to the two opposite boundaries of the domain in such a way that the potential difference was applied either along axis $x$ or along axis $y$.

\section{Results\label{sec:results}}

\begin{figure}[!htb]
  \centering
  \includegraphics[width=\columnwidth]{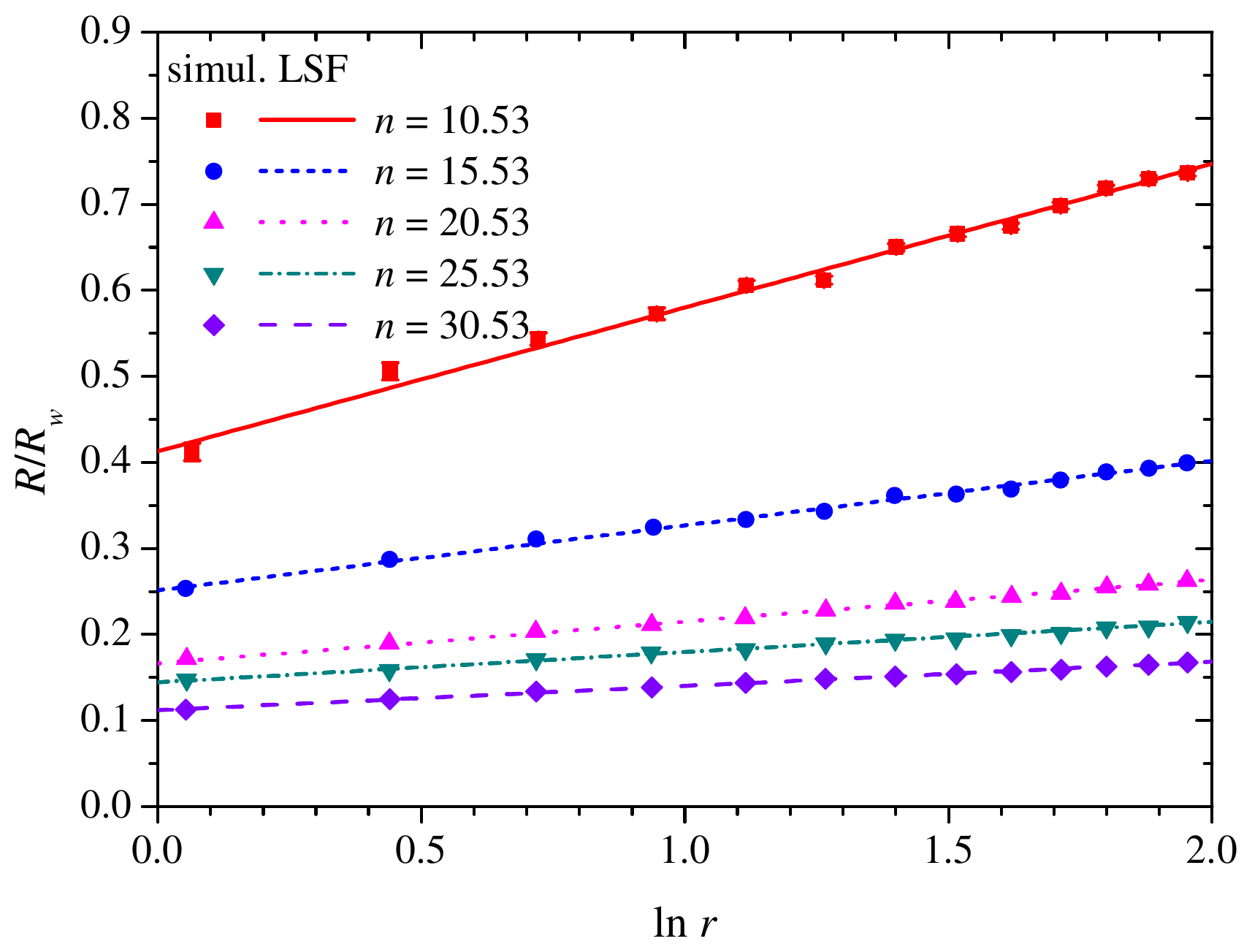}
  \caption{Resistance between two nodes against the distance between these nodes.  Direct computations for RNNs hawing different number density of wires when the junction resistance is negligible as compared to the wire resistance. Lines correspond to the least squares fitting.}\label{fig:R2WDR}
\end{figure}

Figure~\ref{fig:R2WDR}  demonstrates the behavior of the  resistance between two points in RNNs with different values of the number densities of nanowires, when $R_w \gg R_j$. Figure~\ref{fig:R2WDR} reveals that the resistance between two points is proportional to the logarithm of the distance between these two points when the number density of wires is sufficiently large ($n \gtrapprox 2n_c$).

\begin{figure}[!htb]
  \centering
  \includegraphics[width=\columnwidth]{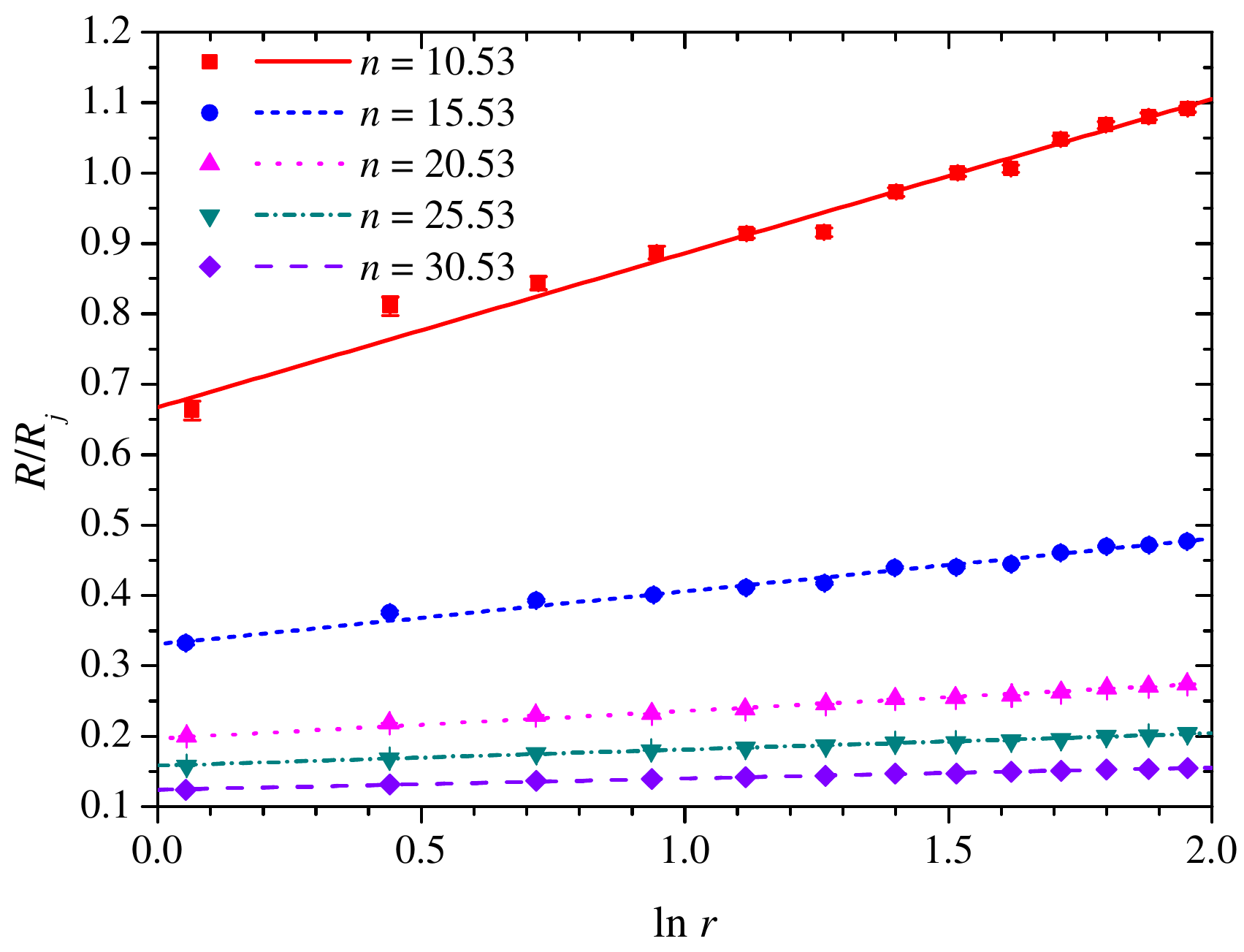}
  \caption{Resistance between two nodes against the distance between these nodes.  Direct computations for RNNs having different number density of wires, when the junction resistance dominates over the wire resistance. Lines correspond to the least squares fitting.}\label{fig:R2JDR}
\end{figure}

Figure~\ref{fig:R2JDR}  demonstrates the behavior of the  resistance between two points in RNNs with different values of the number densities of nanowires, when $R_w \ll R_j$. Despite the network is not planar in this case, the resistance between two points is also proportional to the logarithm of the distance between these two points when the number density of wires is sufficiently large.

\begin{figure}[!htb]
  \centering
  \includegraphics[width=\columnwidth]{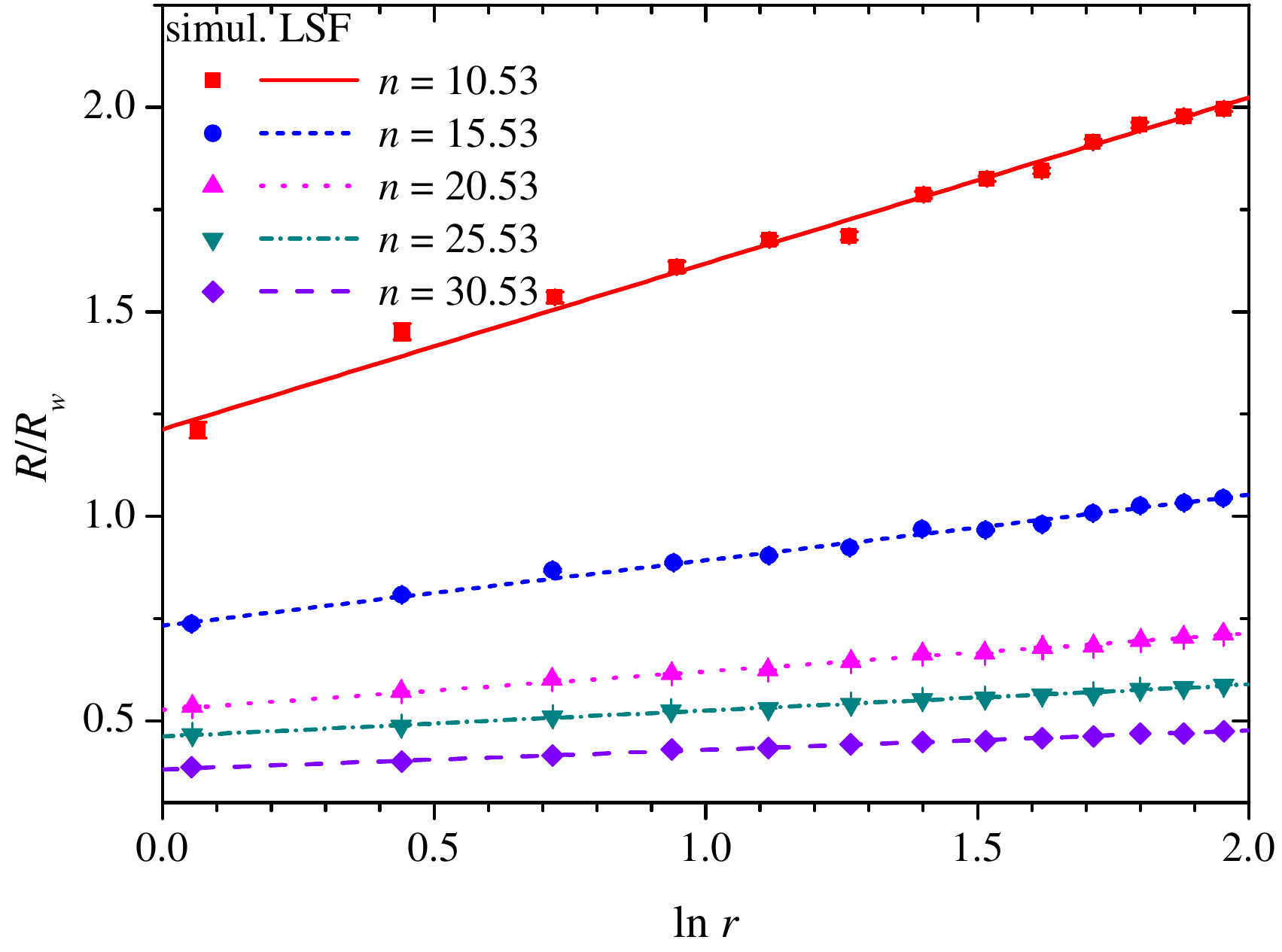}
  \caption{Resistance between two nodes against the distance between these nodes.  Direct computations for RNNs having different number density of wires, when the junction resistance and the wire resistance are equal. Lines correspond to the least squares fitting.}\label{fig:R2JWR}
\end{figure}

Figure~\ref{fig:R2JWR}  demonstrates the behavior of the  resistance between two points in RNNs with different values of the number densities of nanowires, when $R_w = R_j$. Despite the network is not planar in this case, the resistance between two points is also proportional to the logarithm of the distance between these two points when the number density is sufficiently large.

Table~\ref{tab:Rsheetcompare} compares the sheet resistances obtained by means of direct computations using a bus-bar geometry and extracted from Figs.~\ref{fig:R2WDR},\ref{fig:R2JDR}, and \ref{fig:R2JWR} as the slope multiplied by factor~$\pi$.
\begin{table*}[!htb]
  \caption{Comparison of the sheet resistances extracted from the two-point resistance computations and obtained by means of direct computations.\label{tab:Rsheetcompare}}
  \centering
  \begin{ruledtabular}
  \begin{tabular}{ccccccc}
  & \multicolumn{2}{c}{$R_w \gg R_j$}& \multicolumn{2}{c}{$R_w \ll R_j$} & \multicolumn{2}{c}{$R_w = R_j$} \\
  $n-n_c$ & $R_\Box/R_w$, 2-point  & $R_\Box/R_w$, direct & $R_\Box/R_j$, 2-point  & $R_\Box/R_j$, direct &$R_\Box/R_w$, 2-point  & $R_\Box/R_w$, direct \\
  \hline
5 & $0.513 \pm 0.040$ & $0.515 \pm 0.021$    & $0.69 \pm 0.04$ & $0.69 \pm 0.02$         & $1.27 \pm 0.07$ & $1.26 \pm 0.04$ \\
10 & $0.231 \pm 0.008$ & $0.235 \pm 0.010$   & $0.23 \pm 0.02$ & $0.242 \pm 0.004$       & $0.49 \pm 0.04$ & $0.504 \pm 0.007$ \\
15 & $0.147 \pm 0.003$ & $0.151 \pm 0.004$   & $0.119 \pm 0.007$ & $0.1247 \pm 0.0014$   & $0.291 \pm 0.014$ & $0.2922 \pm 0.0025$ \\
20 & $0.108 \pm 0.002$ & $0.111 \pm 0.004$   & $0.072 \pm 0.002$ & $0.0768 \pm 0.0007$   & $0.198 \pm 0.006$ & $0.1995 \pm 0.0014$ \\
25 & $0.088 \pm 0.002$ & $0.0850 \pm 0.0004$ & $0.0496 \pm 0.0012$ & $0.0522 \pm 0.0004$ & $0.150 \pm 0.005$ & $0.1490 \pm 0.0007$ \\
\end{tabular}
\end{ruledtabular}
\end{table*}

\section{Conclusion\label{sec:concl}}

By means of computer simulation, we confirmed the logarithmic dependence of the electrical resistance between the arbitrary nodes of random two-dimensional systems of straight nanowires on the distance between these nodes. In our computations, both the resistance of the conductors and the resistance of the contacts between them were taken into account. The number density of nanowires was well above the percolation threshold, thus almost all nanowires participated in the electrical conductivity. A conductive film based on randomly placed metallic nanowires is a conductive plane. According to\citet{Venezian1994}, the lumped-parameter model of a conductive plane with a sheet resistance equal to $R_\Box$ is a square lattice (grid) of equal resistors of resistance $R_\Box$. Thus, a random nanowire network can be mapped to a square network \cite{OCallaghan2016,He2018,Zeng2022}. However, such a mapping could hardly be produced in rigorous and closed way.

\begin{acknowledgments}
Y.Y.T. acknowledges partial funding from the FAPERJ, Grants No.~E-26/202.666/2023 and No.~E-26/210.303/2023 during his stay in Instituto de F\'{\i}sica, Universidade Federal Fluminense, Niter\'{o}i, RJ, Brasil. 
\end{acknowledgments}

\bibliography{EMTarticle,networks}

\end{document}